\documentclass[%
 reprint,
 superscriptaddress,
 amsmath,amssymb,
 aps,
 apl,
 prb,
]{revtex4-2}

\usepackage{float}
\usepackage{subcaption}
\restylefloat{figure}
\usepackage{graphicx}% Include figure files
\usepackage{dcolumn}% Align table columns on decimal point
\usepackage{bm}% bold math
\usepackage{inputenc}
\usepackage{multirow}
\usepackage{upgreek} %to avoid to have italic greek letters. Now use \upletter. ex. \Uprho \uprho
\usepackage{hyperref}% add hypertext capabilities
\hypersetup{
colorlinks=true,% false: boxed links; true: colored links
linkcolor=blue,% color of internal links
citecolor=blue,% color of links to bibliography
urlcolor=blue% color of external links
}

\graphicspath{{./images/}}

\usepackage{etoolbox}

\makeatletter
\patchcmd{\frontmatter@abstract@produce}
  {\vskip200\p@\@plus1fil
   \penalty-200\relax
   \vskip-200\p@\@plus-1fil}
  {}
  {}
  {}
\makeatother

\begin{document}

\title{A transferable prediction model of molecular adsorption on metals based on adsorbate and substrate properties}% Force line breaks with \\

\author{Paolo Restuccia}
    \email[]{p.restuccia@imperial.ac.uk}
    \affiliation{Department of Chemistry, Imperial College London, 82 Wood Lane, W12 0BZ London, UK}
\author{Ehsan A. Ahmad}
    \affiliation{Faculty of Engineering and the Environment, University of Southampton, University Road, Southampton SO17 1BJ, UK}
\author{Nicholas M. Harrison}
    \affiliation{Department of Chemistry, Imperial College London, 82 Wood Lane, W12 0BZ London, UK}

\begin{abstract}
Surface adsorption is one of the fundamental processes in numerous fields, including catalysis, environment, energy and medicine. The development of an adsorption model which provides an effective prediction of binding energy in minutes has been a long term goal in surface and interface science. The solution has been elusive as identifying the intrinsic determinants of the adsorption energy for various compositions, structures and environments is non-trivial. We introduce a new and flexible model for predicting adsorption energies to metal substrates. The model is based on easily computed, intrinsic properties of the substrate and adsorbate. It is parameterised using machine learning based on first-principles calculations of probe molecules (e.g., H$_2$O, CO$_2$, O$_2$, N$_2$) adsorbed
to a range of pure metal substrates. The model predicts the computed dissociative adsorption energy to metal surfaces with a correlation coefficient of 0.93 and a mean absolute error of 0.77 eV for the large database of molecular adsorption energies provided by \textit{Catalysis-Hub.org} which have a range of 15 eV. As the model is based on pre-computed quantities it provides near-instantaneous estimates of adsorption energies and it is sufficiently accurate to eliminate around 90\% of candidates in screening study of new adsorbates. The model, therefore, significantly enhances current efforts to identify new molecular coatings in many applied research fields.
\end{abstract}

\maketitle

%\tableofcontents

\section*{\label{sec:intro}Introduction}

Control of the chemical and physical properties of surfaces and interfaces is both of fundamental interest and vital in a very wide range of technologies including catalysis, environment, energy and medicine. Given the current pressing need for innovation in areas such as energy supply, energy distribution and transport  there is a pressing need to develop technologies that can both extend the working lifetime of existing
infrastructure and facilitate the development of new sustainable approaches to production and consumption. This has lead to the wide acknowledgment of the importance of controlling material surfaces and coatings~\cite{Barth2005,Choy2003}.
Common phenomena, such as corrosion and friction, cause substantial economic losses every year and
severely impact the environment. For example, in extending the lifetime of current infrastructure, the worldwide costs of prevention, detection and mitigation 
of metal corrosion alone are estimated to be 2.5 trillion US dollars per year~\cite{Koch2016}. In addition, when considering the innovation of new devices, the development of micro- and
nano-electromechanical systems requires new approaches for friction reduction in limited
dimensions which leads to reduced efficiency and failure~\cite{Komvopoulos1996}. The ability to
deposit molecular and nanostructured coatings with advanced functional properties is primed to have a
profound effect on such diverse technologies as wearable electronics, corrosion inhibitors and lubricant additives. One of the challenges in molecular science is therefore the need to find novel, earth-abundant, inexpensive and environmentally friendly materials that adsorb in a controlled manner to surfaces and interfaces.
Historically, innovation of new materials has been a time-consuming and challenging task; it typically takes 20 to 70 years to progress from laboratory conception to widespread commercial use~\cite{Gross2018}.
Developments have also been mainly based on the incremental evolution of existing systems with the oft reported outcome that newly discovered
solutions are based on exactly the same underlying mechanisms as their predecessors; to find something radical and innovative
has usually been a matter of luck.

Extensive use is made of molecular additives for friction and corrosion reduction. A fundamental step in discovering new classes of these surfaces modifiers is a predictive understanding of the thermodynamics for both molecular and
dissociative adsorption on different substrates~\cite{Finsgar2014,Zhu2017,Kousar2021,Neville2007,Minami2017,Fatti2018,Peeters2019}.
For this purpose, it is essential to be able to compute the binding energy (BE) of different adsorption modes with sufficient accuracy to be able to predict the molecular level adhesion of the self assembling coating. In principle this is achievable using modern atomistic simulations but in practice is problematic as the parameter space of factors that affect the BE is very large~\cite{Norskov2008}. There has therefore been a significant and sustained effort aimed at identifying a small number of easily computed descriptors that can accurately capture the nature of the
molecule-surface interaction, and thus facilitating a simple and efficient predictive model of adsorption. 
In recent
years the combination of high-throughput density functional theory (DFT) calculations and
machine learning techniques has opened a new era of informatics-based approach to materials design, from which a number of simple models for predicting adsorption energies have emerged~\cite{Ras2013,Gao2020,Dean2019,Roling2018,Roling2017,Yan2018,Liu2020,Calle-Vallejo2014,Winther2019,Mamun2019,Mamun2020,Andersen2019,Chowdhury2018,Praveen2020,Zhang2021,Chang2021,Fung2021,Li2021,Anderson2020}. Similar approaches have also been used for predicting other figures of merit, like the inhibition efficiency of molecules~\cite{Cher2020}.
These models are usually based on linear relationships using simple descriptors for both
the substrate and the adsorbed molecule (e.g., the number of valence electrons, the electronegativity of the substrate~\cite{Gao2020} and the ionization potential of the molecule~\cite{Dean2019}). Despite their simplicity, these models have been demonstrated to be quite effective in
predicting adsorption energies, especially when machine learning techniques are employed~\cite{Ras2013,Dean2019}. However, in the past such models have been limited in their
transferability. For instance, any given model may be limited to the adsorption of molecules in
one specific adsorption site (i.e., on-top or hollow). For a more extensive employment of these
predictive calculations, one would like to extend the possible range of adsorption sites to
model a broader array of realistic configurations, such as stepped edges or grain boundaries
at the surface, and include a wide range of molecular adsorbates.

In the current work we present a new predictive linear model that uses appropriate physical
descriptors to predict in minutes the adsorption of a wide range of molecules to multiple substrates in a variety
of surface adsorption sites. The model is based on a combination of systematic DFT calculations and machine learning.
The model reported here accurately predicts the different dissociative adsorption energies of a range of probe
molecules over simple homogeneous metallic substrates. Despite its simplicity, the model
provides a good estimate of molecular BE in different configuration sites with limited computational effort, and it is devised in a form
that facilitates its extension to more complicated structures (e.g., oxides, carbonates or defective surfaces).
Moreover, the margin of error in the BE prediction is sufficiently small that provides a sufficiently accurate
estimate of fully optimised \textit{ab initio} calculations, saving time and facilitating the rapid screening
of a broader range of systems. Therefore, the model is accurate enough to guide the discovery and optimisation of
molecular adsorbates in order to improve the functionality of corrosion inhibitors and lubricant additives and is
likely to find application in fields such as catalysis, molecular electronics and
biomedicine~\cite{Bligaard2004,Joachim2005,Kasemo2002}, where the adsorption of molecules and molecular
films underpins many important processes.

\section*{\label{sec:method}Method}

\subsection*{\label{sec:slab_calc}BE calculation in slab configuration}

For the adsorption energies of the training set (i.e., the slab systems) and the calculation of the
later defined Molecule-Bulk Energy (MBE) terms, Spin-polarized DFT calculations were performed using the
projector-augmented wave method (PAW) as implemented in the plane-wave code QUANTUM ESPRESSO (QE)~\cite{QE}.
We used the PAW pseudpopotentials~\cite{Kresse1999} from the PSLibrary 1.0.0~\cite{DalCorso2014} within the
generalised gradient approximation (GGA) of Perdew, Burke and Enzerhoff (PBE)~\cite{PBE1996} for the
exchange-correlation energy. The electronic wave functions are expanded as a linear combination of
plane waves up to a kinetic energy of 95 Ry, which we find is sufficient to converge the total energies
(1 meV/atom) and equilibrium lattice constant (0.1 m\r{A}) for the considered substrates. 

For the cell structure, we considered different configurations for the slab and the MBE
calculations: for the former, we employed supercells with a $2 \times 2$ in-plane size in order to
reduce the interaction between adsorbate replicas and 5 of 6 layers, depending of the substrate of
the different materials. For the latter, all the clusters were computed in a $20 \: \text{\r{A}}
\times 20 \: \text{\r{A}} \times 20 \: \text{\r{A}}$ cubic cell, so the self interaction between the cluster
replicas is negligible. All the input and output geometries for both the slab and the MBE calculations
are provided as Supplementary Information.

The Monkhorst-Pack grid~\cite{MPgrid} is used for sampling the Brillouin Zone, but different $k$-mesh
for each structure under study were considered. In particular, we selected the optimal $k$-point grid
for each slab geometry, whereas all the calculations involving clusters had a sampling at Gamma point
due to the large cell dimensions. To improve the convergence, the Marzari-Vanderbilt cold
smearing~\cite{Marzari1999} method is used for the sampling of the Fermi surface, with a width of
0.27 eV in order to obtain accurate forces. The convergence criteria of forces and energy are 0.003
eV \r{A}$^{-1}$ and 10$^{-2}$ eV.

\subsection*{\label{sec:comp_appr}HOMO, LUMO and HOMO-LUMO gap calculation for molecule in gas phase}

For the calculation of the HOMO, LUMO and HOMO-LUMO gap, we performed DFT calculations using the CRYSTAL17 computational
suite~\cite{Crystal17,Crystal17-man}, in which the crystalline orbitals are expanded as a linear combination of a
local basis set composed by atom-centered Gaussian orbitals with s, p, or d symmetry. For all the elements employed
in the molecular calculations (namely, H, C, N, O, F, S, Cl), we used the 6-31G** basis
sets~\cite{Clark1983,Ditchfield1971,Francl1982,Gordon1982,Hariharan1973,Hehre1972,Spitznagel1987}.

The approximation of the exchange and correlation functional is based on the Becke, 3-parameter,
Lee-Yang-Parr (B3LYP) hybrid functional incorporating 20\% Hartree-Fock
exchange~\cite{Becke1993,Lee1988,Stephens1994}. The Coulomb and exchange series are summed directly and
truncated using overlap criteria with thresholds of 10$^{-10}$, 10$^{-10}$, 10$^{-10}$, 10$^{-20}$,
10$^{-30}$ as described elsewhere~\cite{Crystal17-man,Pisani1988}.

\section*{\label{sec:results} Results and Discussion}

The general definition for the computed BE to a surface may be written as:

\begin{equation}
    \text{BE} = \text{E}_{\text{tot}} - \left( \text{E}_{\text{sub}} + \text{E}_{\text{mol}} \right)
\end{equation}
\\
where $\text{E}_{\text{tot}}$ is the computed total energy for a system composed of a molecule adsorbed
on a substrate and $\text{E}_{\text{sub}}$ ($\text{E}_{\text{mol}}$) is the energy for the isolated
substrate (molecule). With this definition, a negative (positive) BE indicates that the dissociation process is favourable (unfavorable).

The total BE can be analysed in terms of many contributions that may be related to properties of the molecule and the surface \cite{Scaranto2011,Morin2004,Wang2010}. Defining a comprehensive model for the BE is challenging. A recent approach proposed by Dean \textit{et al.} \cite{Dean2019} succeeded in predicting the BE of probe molecules to metal nano-particles. This model is based on the idea that
BE can be adequately represented by stability descriptors for the adsorbate, the adsorption site, the substrate and a simply computed estimate of the interaction between the molecule and the surface. These assumptions led to the
following linear equation for the BE:

\begin{equation}
    \text{BE} = a + b \times \text{CE}_{\text{local}} + c \times \text{IPEA} + d \times \text{MADs}
    \label{eq:sci_paper}
\end{equation}
\\
where $\text{CE}_{\text{local}}$ is the term to describe the local cohesive energy of the adsorption
site, $\text{IPEA}$ is the negative average between the ionization potential and the electronic affinity
of the molecule, and the $\text{MADs}$ is the gas phase BE between the adsorbate and one atom of the
metal substrate, which is obtained through \textit{ab initio} calculations, and represents the descriptor
for the adsorbate-metal interaction. Although this model proved to be effective in the prediction of BE,
with correlation coefficient $R^2$ of around 0.94 and a mean absolute error (MAE) of around 0.1 eV, there
are some limitations in the employed approach: i) the adsorbates were always in an on-top site
configuration, limiting the possibility to predict the BE in other adsorption sites such as hollow or
bridge, and ii) the model has been trained only on noble metals nano-particles and slab surfaces, such
as Ag, Au and Cu, narrowing the range of possible substrates over which the prediction is effective.

In order to overcome these limitations, we present here a model using suitable descriptors for the adsorption of molecules over flat substrates. In particular, we propose the following equation for the
prediction of BE:

\begin{equation}
    \text{BE} = a + b \times \text{CE}_{\text{B}} + c \times \left( \text{W}_{\text{F}} - \frac{ \text{E}_{\text{gap}}}{2} \right) + d \times \text{MBE}
    \label{eq:model}
\end{equation}
\\
where $\text{CE}_{\text{B}}$ is the cohesive bulk energy for the substrate atomic species,
$\text{E}_{\text{gap}}$ is the gap between the Highest Occupied (HOMO) and Lowest Unoccupied (LUMO) Molecular Orbital of the adsorbed molecule (from now on, HOMO-LUMO gap), $\text{W}_{\text{F}}$
is the work function of the substrate, $\text{MBE}$ is the Molecule-Bulk Energy, which resembles
the $\text{MADs}$ of Eq.~\ref{eq:sci_paper} and it is computed using \textit{ab initio} theory; $a$, $b$, $c$ and $d$ are the linear coefficients for the regression.
$\text{CE}_{\text{B}}$ provides a general estimate of the strength of the interaction between the substrate atoms and the $\text{MBE}$ provides a simply computed estimate of the substrate-molecule interaction. The third term contains the difference between the surface work function and the middle of the HOMO-LUMO gap
of the adsorbed molecule which in frontier molecular orbital theory controls the charge transfer and hybridisation contributions to the surface binding~\cite{Fukui1952,Hoffmann1988,Ishii1999}. The $\text{MBE}$ term is computed as:

\begin{equation}
    \text{MBE} = \sum_{i = 1}^{n_{frag}} E_{complex, i} - E_{B, M, i} - \mu_{G, mol, i}
    \label{eq:mbe_mini}
\end{equation}
\\
where $n_{frag}$ is the number of molecular fragments considered in the dissociative adsorption process,
$E_{complex}$ is the total energy of a molecular functional group adsorbed on a single atom of the metal
substrate, $E_{B, M}$ is the bulk energy of a single atom of the substrate atomic species and $\mu_{G,
mol}$ is the chemical potential of the molecular fragment generalised from the fragment energy to allow for the adsorption environment.
This quantity provides an easily computed and flexible estimate of the strength of adhesion between the adsorbate and the substrate.
In contrast to the MADs term proposed by Dean \textit{et al.}, where all the functional groups are computed as isolated components, in the proposed MBE term we refer all energies to a consistent reference enabling the use of pre-computed data in a transferable predictive model.
Another advantage of this approach is choosing the proper reference for the chemical potential
in the calculation of MBE. In the current work, we chose to refer $\mu_{G, mol}$ to the isolated gas phase molecule for the sake of simplicity. However, it is possible to reference the chemical potential to different environments including solvated species, as shown in recent electrochemical
studies~\cite{Hormann2019,Hormann2020,Hormann2021}.

\begin{figure}[htpb]
    \centering
    \includegraphics[width=0.9\linewidth]{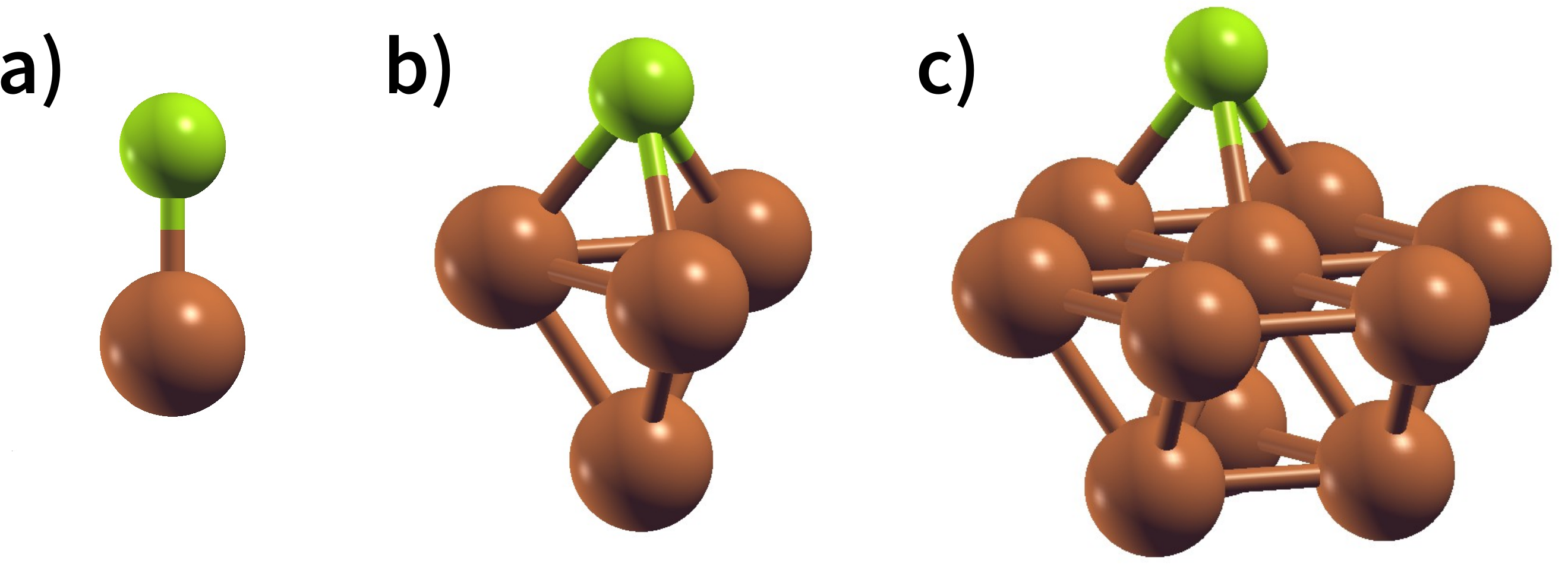}
    \caption{Ball and stick representation of the models used for the different approaches in the calculation of $\text{MBE}$ in the case of Cl adsorption on Cu: a) the substrate is modelled by just one atom, b) the substrate is represented by a cluster of 4 atoms and c) the cluster modelling the substrate is composed by 10 atoms. Green and brown balls represent chlorine and copper atoms, respectively.}
    \label{fig:mbe_struct}
\end{figure}

In the current work, ordinary least squares (OLS) linear regressions were used to determine the coefficients in Eq.~\ref{eq:model} from a training set of {\em ab initio} energies 
using the \textit{statsmodels} library~\cite{Seabold2010} provided in Python 3~\cite{Python3}. For the OLS
regression, we adopted a training set of eight different probe molecules, namely Cl$_2$,
CO$_2$, F$_2$, H$_2$, H$_2$O, H$_2$S, N$_2$ and O$_2$, adsorbed over ten different metal substrates,
namely Ag(111), Al(111), Au(111), Cu(111), Fe(100), Fe(110), Ir(111), Pt(111), V(100) and V(110). In each case the energy of the most stable adsorption configuration was used. Where possible standard reference data was used for each of the terms of Eq.~\ref{eq:model}: for $\text{CE}_{\text{B}}$, we used the observed formation energies of the transition metals provided by Ref.~\cite{Kittel2004}, for the work function, we employed the DFT computed values provided by the Materials Project database~\cite{Jain2013}. The HOMO-LUMO gap of the molecules was estimated from 
\textit{ab initio} calculations, with the computational details provided in the
Methods section. The MBE was also computed \textit{ab initio} using a small cluster which will be discussed below.

\subsection*{Analysing Contributions to the MBE}

\begin{table}[t]
\caption{Regression coefficients, i.e., Coefficient Estimate, Standard Error (SE) and P-value, for the different approaches employed for MBE calculation in the case of a) single metal atom, b) a small metal cluster and c) a large metal cluster. Cases are trained using the dataset provided in the Supplementary Information. $R^2$ is the correlation coefficient, MAE is the mean absolute error.}\label{tab:regr_coeff}
\centering
\begin{subtable}{\linewidth}
\caption{First approach for MBE, as shown in Fig.~\ref{fig:mbe_struct}a. $R^2 = 0.21$, $\text{MAE} = 1.97$ eV, $\text{RMSE} = 2.52$ eV.}\label{tab:regr_coeff-a}
\centering
{
\begin{tabular}{c c c c}
\hline
  & Coefficient Estimate & SE & P-value \\
\hline 
$a$ &  2.2460 & 1.6048 & 0.167 \\
$b$ & -0.7268 & 0.3114 & 0.023 \\
$c$ &  0.8900 & 0.2721 & 0.002 \\
$d$ & -0.0886 & 0.0970 & 0.365 \\
\hline
\end{tabular}
}
\end{subtable}%

\vspace*{0.75 cm}

\begin{subtable}{\linewidth}
\caption{Second approach for MBE, as shown in Fig.~\ref{fig:mbe_struct}b. $R^2 = 0.83$, $\text{MAE} = 0.89$ eV, $\text{RMSE} = 1.17$ eV.}\label{tab:regr_coeff-b}
\centering
{
\begin{tabular}{c c c c}
\hline
  & Coefficient Estimate & SE & P-value \\
\hline 
$a$ &  1.5812 & 0.7064 & 0.029 \\
$b$ & -0.2925 & 0.1461 & 0.050 \\
$c$ &  0.1793 & 0.1122 & 0.116 \\
$d$ &  1.0163 & 0.0691 & $3 \cdot 10^{-21}$ \\
\hline
\end{tabular}
}
\end{subtable}%

\vspace*{0.75 cm}

\begin{subtable}{\linewidth}
\caption{Third approach for MBE, as shown in Fig.~\ref{fig:mbe_struct}c. $R^2 = 0.94$, $\text{MAE} = 0.52$ eV, $\text{RMSE} = 0.69$ eV.}\label{tab:regr_coeff-c}
\centering
{
\begin{tabular}{c c c c}
\hline
  & Coefficient Estimate & SE & P-value \\
\hline 
$a$ &  0.7426 & 0.4208 & 0.083 \\
$b$ & -0.1735 & 0.0874 & 0.052 \\
$c$ &  0.1844 & 0.0659 & 0.007 \\
$d$ &  0.9927 & 0.0370 & $3 \cdot 10^{-34}$ \\
\hline
\end{tabular}
}
\end{subtable}%

\end{table}

\begin{figure}[htpb]
    \centering
    \includegraphics[width=0.9\linewidth]{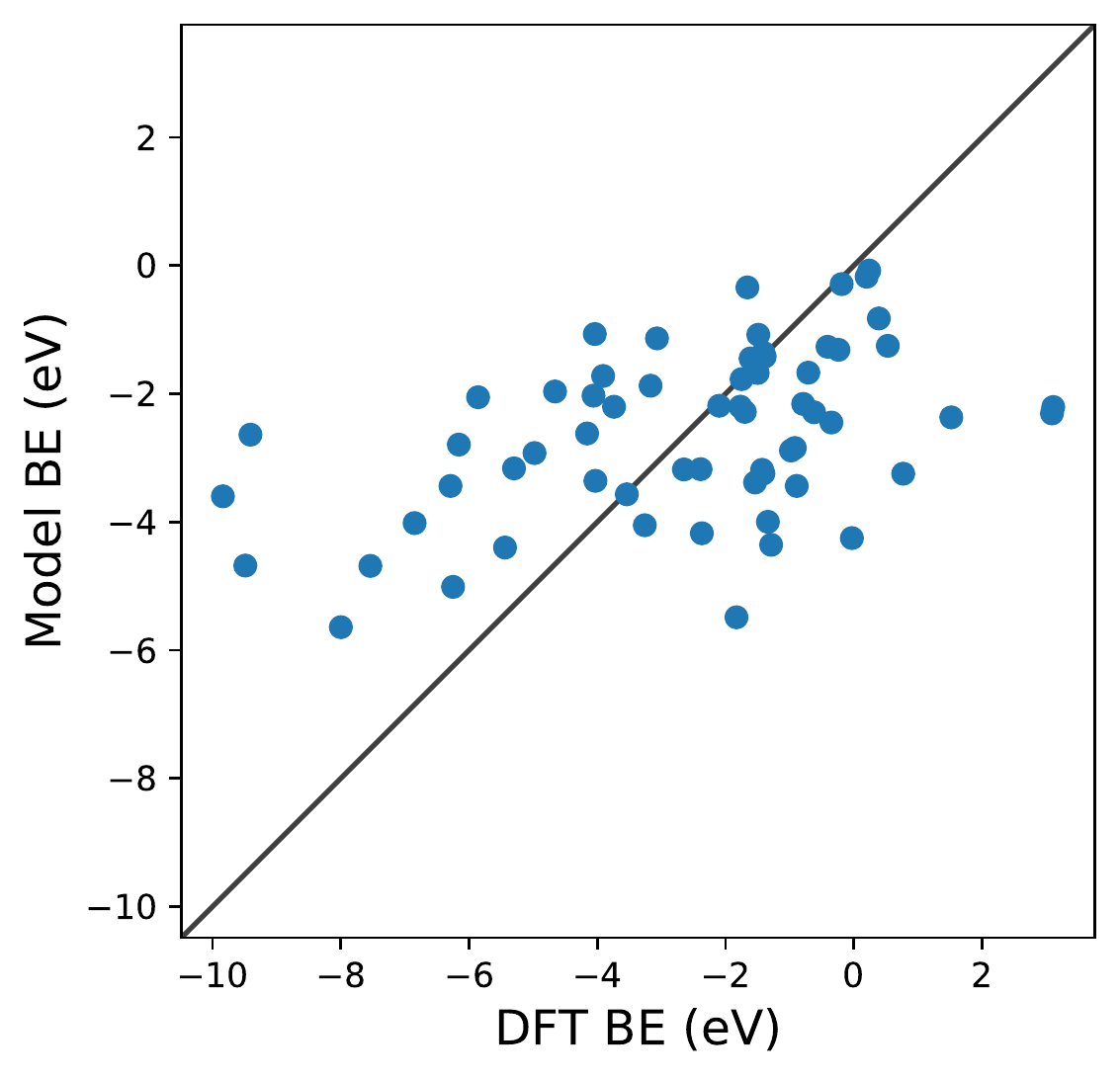}
    \caption{Parity plot for the training of the model against the DFT BE calculations with the MBE approach proposed in Eq.~\ref{eq:mbe_mini} and the system shown in Fig.~\ref{fig:mbe_struct}a. The black solid line represents the parity between the computed DFT BE and the predicted value.}
    \label{fig:mbe_mini}
\end{figure}

An appropriate calculation of the MBE is essential for the efficiency and accuracy of the proposed model. The simplest level of approximation used here is that proposed by Dean {\em et al.} in which the MBE is computed as
Eq.~\ref{eq:mbe_mini}, i.e., the binding energy is the energy difference in the gas phase of a specific
fragment obtained during the dissociative process and one metal atom of the substrate~\cite{Dean2019}. An example of this
possible configuration to calculate MBE is shown in Fig.~\ref{fig:mbe_struct}a for the case of Cl adsorbed to a Cu atom. The regression statistics are shown in Table~\ref{tab:regr_coeff-a}, while Figure~\ref{fig:mbe_mini} shows the parity plot of the 
model training against the DFT computed adsorption energies for the predicted BE. 

This approximation provides a rather poor prediction of the BE: the correlation coefficient $R^2$ is around 0.21 (an $R^2$ of 0 corresponds to no correlation and of 1.0 to a perfect set of predictions) and the mean absolute error (MAE) is almost 2 eV. The parity plot in Figure~\ref{fig:mbe_mini} confirms that the linear regression does not provide a good description of the BE. 
Although Dean \textit{et al.} have shown convincingly that this approach reproduces the energy of adsorption to metallic nanoparticles in an on-top configuration of several radical groups (namely, CH$_3$, CO and OH), it evidently fails to do so when the molecules are adsorbed on a wide range of substrates.

A possible explanation for this discrepancy is that the variations of the interactions in the hollow and bridge adsorption sites considered here are not captured by binding to a single metal atom. This suggests that a somewhat larger cluster is required to take into account the different adsorption site configurations in the calculation of MBE such as that represented in Figure~\ref{fig:mbe_struct}b. Here the Cl is adsorbed to a four atom cluster based on the hollow site presented by the Cu(111). This is the smallest cluster, for this specific substrate, which retains the symmetry of the surface adsorption site. We conveniently create these clusters that resemble the surface adsorption sites for all the considered substrates in our training set and the geometries employed for these calculations are provided as Supplementary Information. Another essential advantage of this approach is the possibility to explore sites with lower coordination numbers that resemble substrates containing defects or stepped edges. The only possibility to simulate these configurations with a periodic slab structure is by using large supercells with hundreds of atoms, thus increasing the computation time significantly compared to the few atoms used within a cluster.

\begin{figure}[h!]
    \centering
    \includegraphics[width=0.9\linewidth]{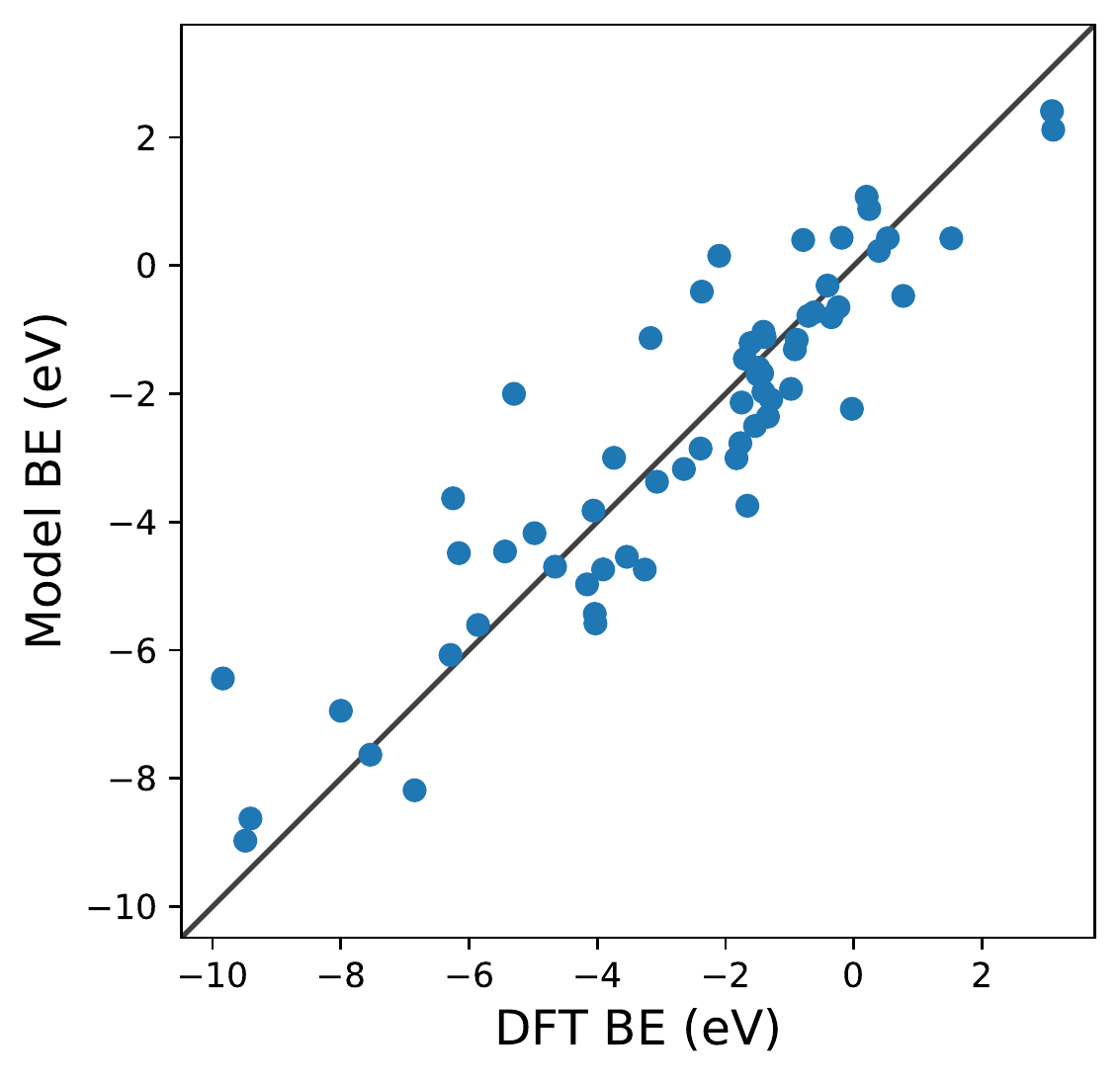}
    \caption{Parity plot for the training of the model against the DFT BE calculations with the MBE cluster approach proposed in Eq.~\ref{eq:mbe_clust} and the system shown in Fig.~\ref{fig:mbe_struct}b. The black solid line represents the parity between the computed DFT BE and the predicted value.}
    \label{fig:mbe_small}
\end{figure}

With the use of this approach, we change the definition of MBE as follows:

\begin{equation}
    \text{MBE} = \sum_{i = 1}^{n_{frag}} E_{complex, i} - E_{cluster, i} - \mu_{G, mol, i}
    \label{eq:mbe_clust}
\end{equation}
\\
where all the terms of Eq.~\ref{eq:mbe_clust} are the same as Eq.~\ref{eq:mbe_mini}, apart from
$E_{cluster, i}$ which is the energy of the cluster modelling the substrate. This approach leads to a significant improvement in the BE prediction, as shown in both
Table~\ref{tab:regr_coeff-b} and Figure~\ref{fig:mbe_small}. There is 
a significant improvement in both the correlation coefficient (around 0.83) and the MAE (around 0.9 eV).

Extending this approach, one can compute the MBE from adsorption to the 10 atom cluster displayed in Figure~\ref{fig:mbe_struct}c for the case Cl adsorbed to a hollow site on Cu. This cluster also maintains the adsorption site symmetry.

\begin{figure}[htpb]
    \centering
    \includegraphics[width=0.9\linewidth]{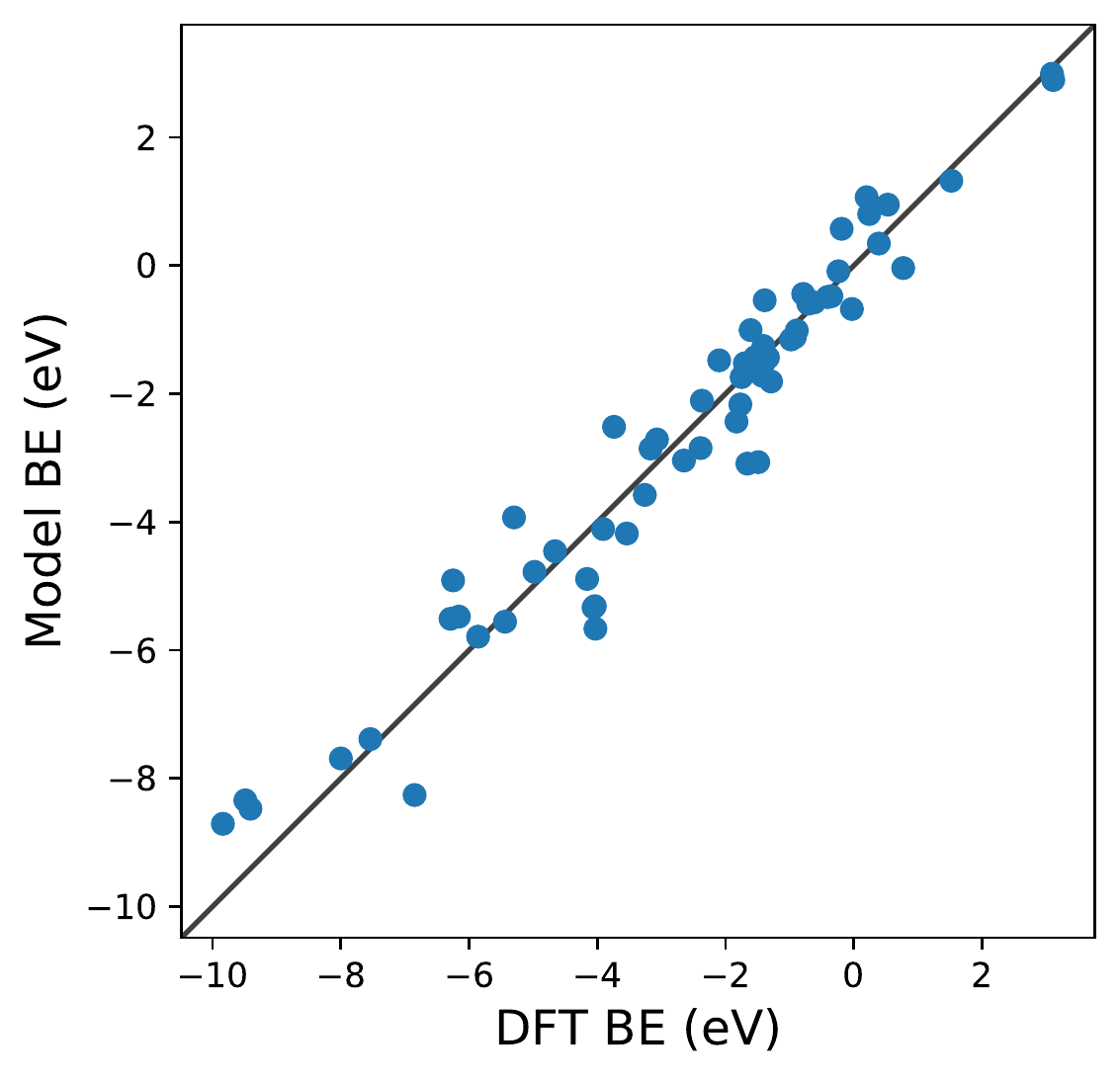}
    \caption{Parity plot for the training of the model against the DFT BE calculations with the MBE cluster approach proposed in Eq.~\ref{eq:mbe_clust} and the system shown in Fig.~\ref{fig:mbe_struct}c. The black solid line represents the parity between the computed DFT BE and the predicted value.}
    \label{fig:mbe_large}
\end{figure}

From Figure~\ref{fig:mbe_large} it is evident that the model based on this MBE provides a satisfactory description of the adsorption energetics with a correlation coefficient of 0.94, and a more significant reduction of the MAE to 0.5 eV.

In addition, all of the fitted parameters have a
P-value equal or smaller than 0.05, which is the threshold to obtain a confidence level of 95\% in the predictions of the model.
Even if this threshold should not be seen as a sharp edge for statistical significance~\cite{AMC2020},
the obtained values for both the P-value and the standard errors provide a rigorous test for the
effectiveness of our model.

\begin{figure*}[htpb]
    \centering
    \includegraphics[width=0.4\linewidth]{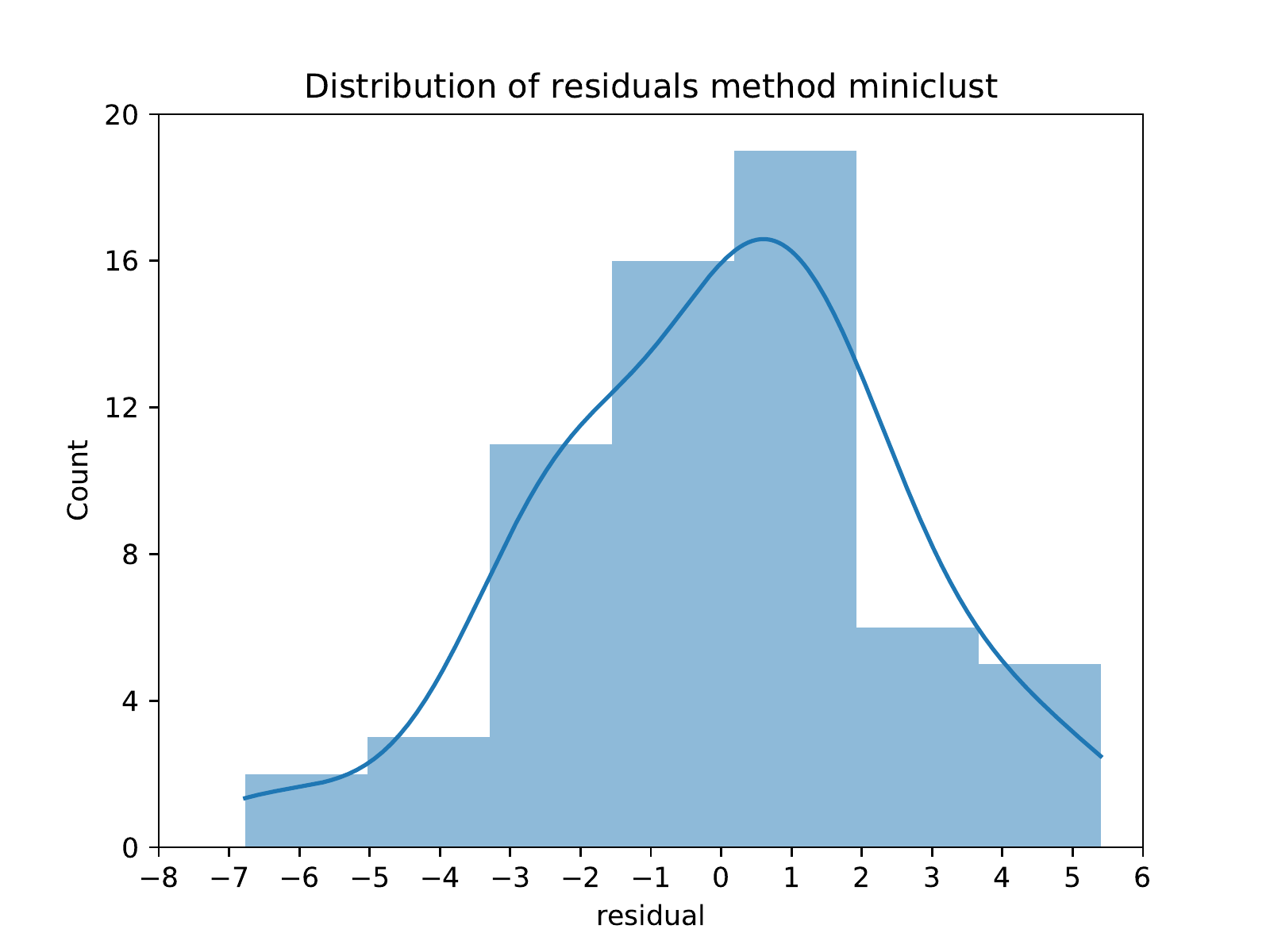}
    \includegraphics[width=0.4\linewidth]{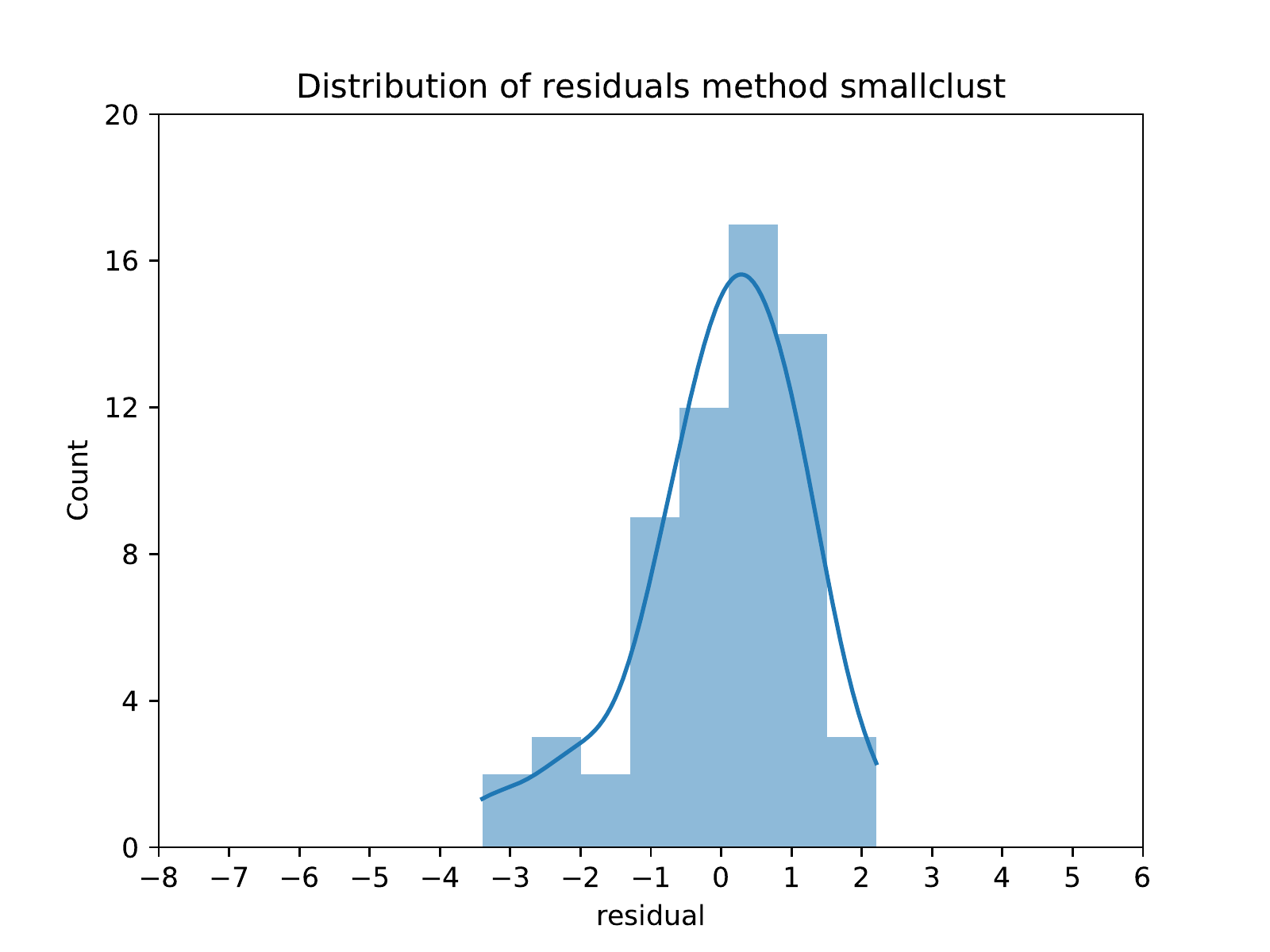}
    \includegraphics[width=0.4\linewidth]{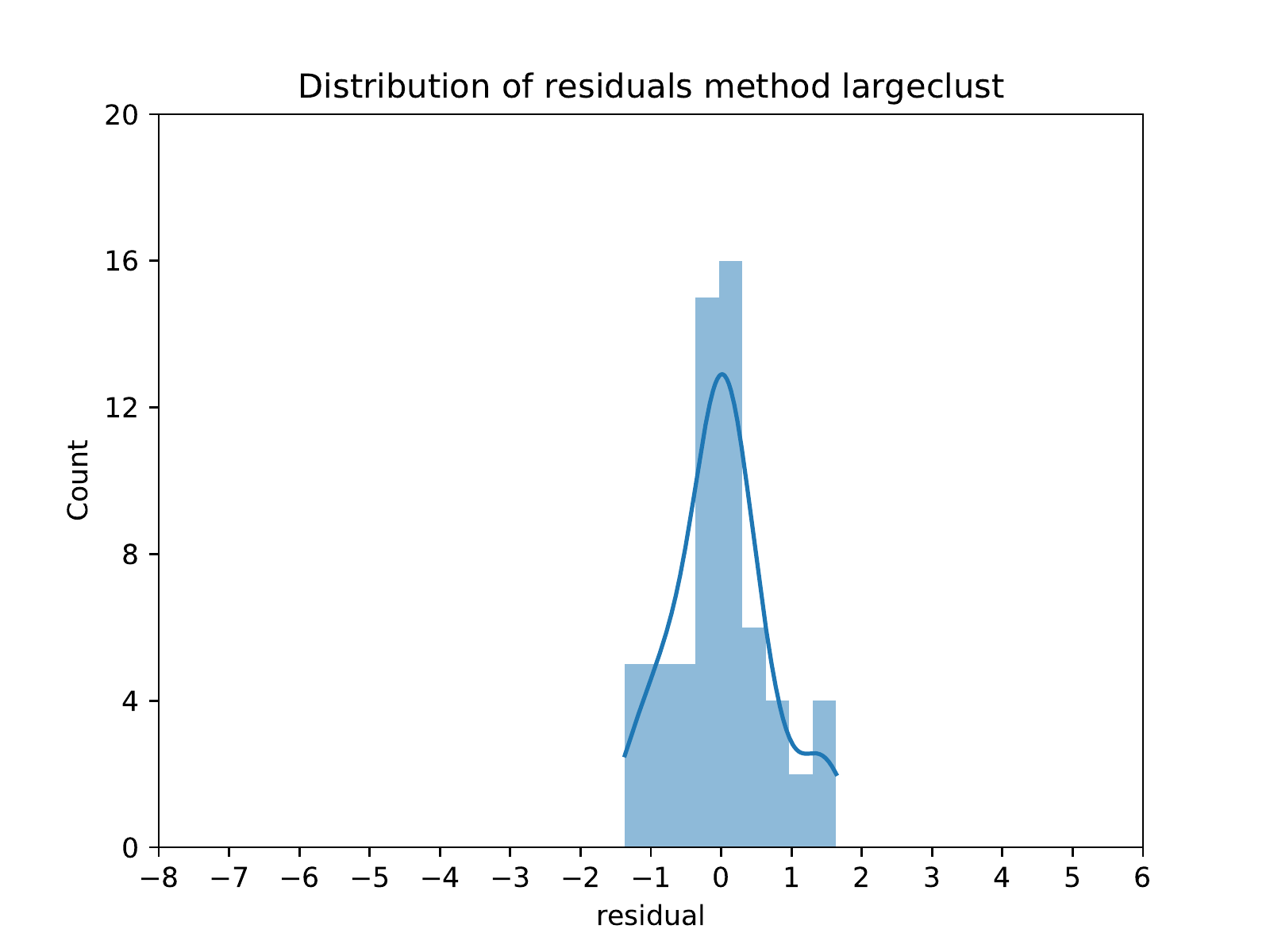}
    \caption{Residual distribution for the MBE calculation as proposed in Fig.~\ref{fig:mbe_struct}a (left panel), Fig.~\ref{fig:mbe_struct}b (center panel) and Fig.~\ref{fig:mbe_struct}c (right panel). The blue curve represents a smoothing interpolation of the residual data.} 
    \label{fig:residuals}
\end{figure*}

It is possible to perform a more qualitative analysis of the accuracy of the proposed models by considering the
residuals distribution as shown in Fig.~\ref{fig:residuals}: each panel presents this distribution as a
histogram counter of the errors for the MBE calculation approaches proposed in this work. As expected,
the errors follow a normal distribution, and every time the description of the MBE is improved, the residuals
range is almost halved, further validating our approach. It is also interesting to notice that moving
from the smaller to larger cluster leads to a notable reduction in the MAE of around 40\%, but the correlation
coefficients are similar (0.83 for the smaller clusters, 0.94 for the larger ones). Therefore, one can adjust
the model for convenience: it is possible to either lose a bit of predictive accuracy with the advantage
of realising a database with smaller clusters (which usually requires around half the time for the actual
calculation) or retain a better precision with the cost of longer computational time to build the database of
the MBE coefficients. From now on, our analysis will focus on the predictive model employing the larger
clusters since they provide the most accurate results.

The model proposed here is therefore able to predict BE with a fidelity comparable to the current state of the art but for a variety of adsorption geometries and adsorbates. The MAE for the training set is somewhat higher that of the model reported by Dean \textit{et al.} of around 0.1 eV but it is computed for a training dataset with a much larger range of BE (from -10 to +3 eV), so the associated relative error is comparable.

\begin{table}[htpb]
\centering
\caption{Correlation coefficient $R^2$, mean absolute error (MAE) and root mean square error (RMSE) of the OLS regression for the proposed Model and the MBE against the computed DFT values for the BE as shown in Figure~\ref{fig:model_vs_mbe}.}\label{tab:model-mbe}
\begin{tabular}{c c c}
\hline
  & Model & MBE \\
\hline 
$R^2$     & 0.94 & 0.93 \\
MAE (eV)  & 0.52 & 0.56 \\
RMSE (eV) & 0.69 & 0.79 \\
\hline
\end{tabular}
\end{table}

\begin{figure}[htpb]
    \centering
    \includegraphics[width=0.9\linewidth]{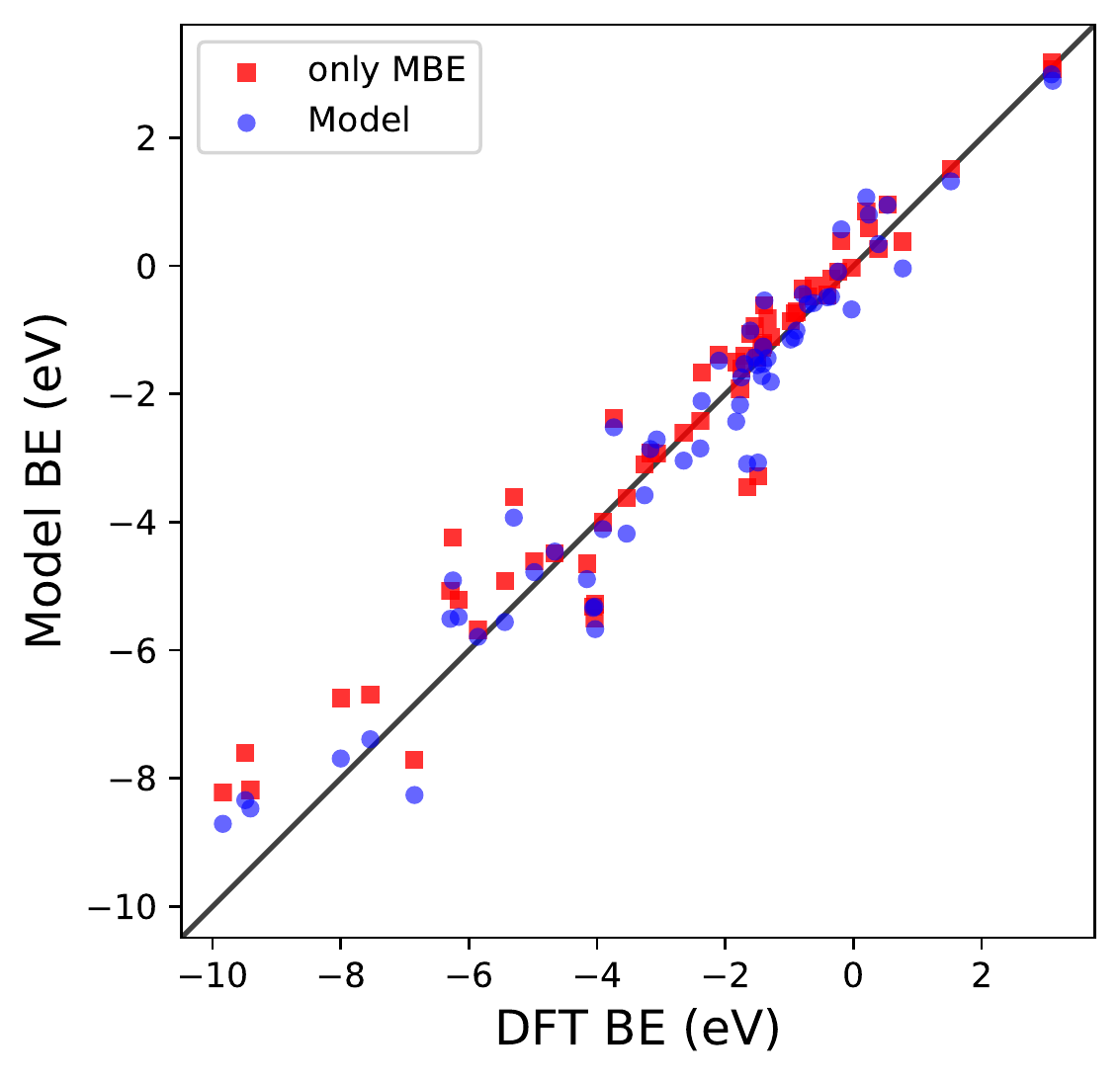}
    \caption{Parity plot for the trained model BE (blue dots) with the coefficients reported in Table~\ref{tab:regr_coeff-c} and MBE values (red squares) against the DFT BE calculations.}
    \label{fig:model_vs_mbe}
\end{figure}

Before proceeding with the model validation analysis, it is interesting to note the
importance of MBE in calculating the BE since its OLS regression coefficient $d$ is the one
with the smallest relative error and P-value. To understand how relevant this term is in calculating
the predicted BE, we compare two different types of dataset training. The first is the one we
discussed in the previous paragraph and is shown in Figure~\ref{fig:mbe_large}. The second is based
on a simple OLS regression of the MBE values of the considered reaction paths against the computed
DFT BE. The results are shown in Figure~\ref{fig:model_vs_mbe} and Table~\ref{tab:model-mbe}.
It is apparent that qualitatively the first training approach (blue dots) provides similar results
to the one based solely on MBE (red squares), highlighting the greater importance of the MBE term in the
definition of the model. A deeper analysis involving the regression coefficients, such as $R^2$, MAE
and the root mean square error (RMSE), shows us a clearer picture. Although $R^2$ is similar in both
scenarios (0.94 vs 0.93), we notice an increase in both MAE and RMSE when considering in training only
MBE by 8\% and 14\%, respectively. Therefore, even if the MBE is an essential part of the definition of
this new model, it is important to consider all the physical terms we have identified in the definition
of Eq.~\ref{eq:model}, in order to minimize the average error in the BE.

After the training, the following step is to validate the model for use in predicting new
dissociation paths for larger molecules. To do so, we compared its predictions to reaction energies
computed and tabulated in previous work.

\subsection*{Validation of the Model}

\begin{figure*}[htpb]
    \centering
    \includegraphics[width=0.43\linewidth]{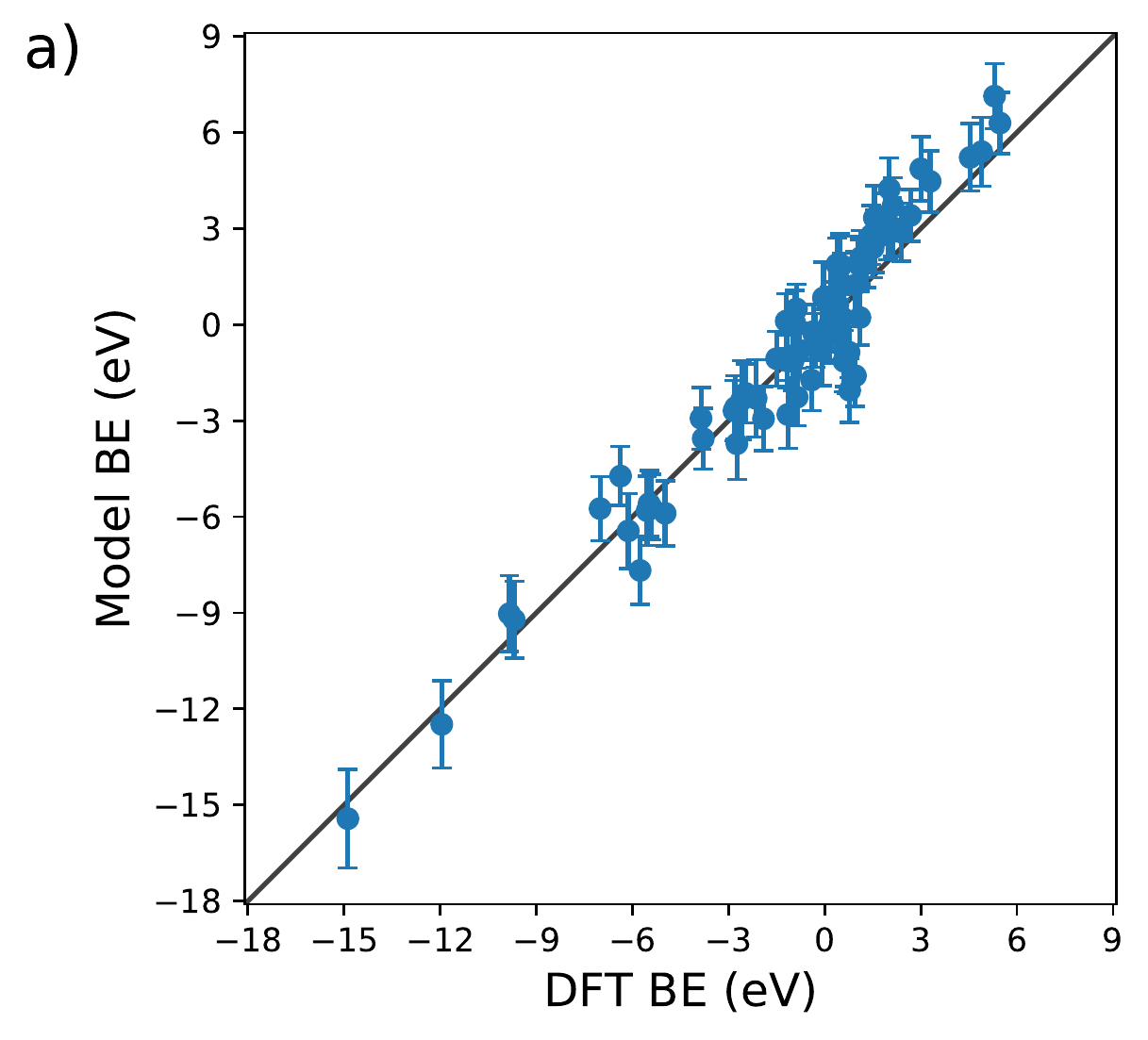}
    \includegraphics[width=0.43\linewidth]{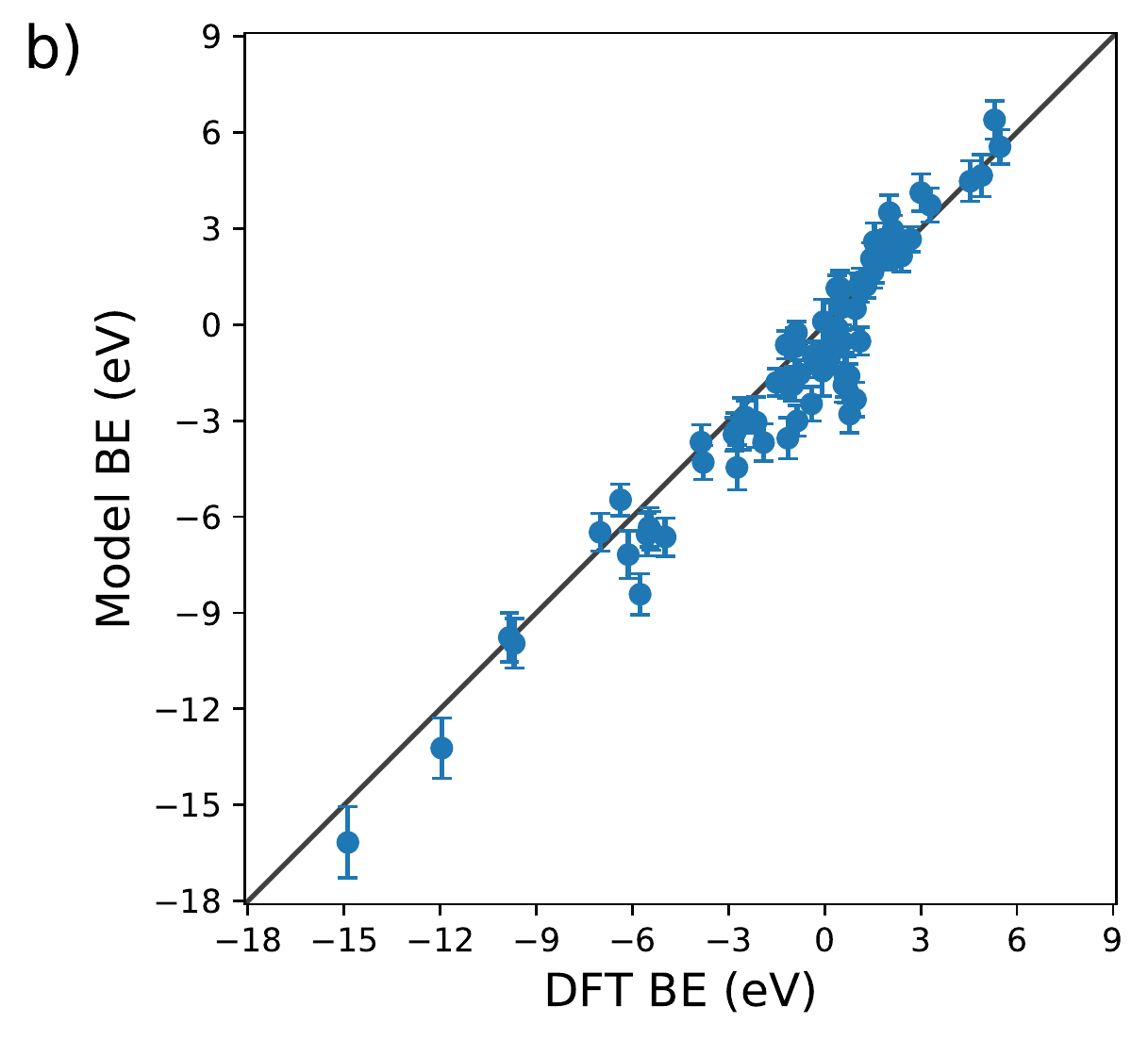}
    \caption{Parity plots for the validation of the model against the DFT BE calculations available on the Catalysis-Hub.org database. The panel a) shows the predicted values considering the intercept $a$, whereas panel b) represents the predictive model dismissing the intercept $a$.}
    \label{fig:mbe_validation}
\end{figure*}

The validation of the proposed model is essential for its employment in actual
technological applications. To do so, we compare the BE predicted by the model with the DFT reaction
energies available on the Catalysis-Hub.org database~\cite{Winther2019,Mamun2019} for 80 different
surface reactions. In particular, we chose twelve different molecules, namely CH$_4$, C$_2$H$_6$, CO$_2$,
H$_2$, H$_2$O, H$_2$S, N$_2$, NH$_3$, NO, N$_2$O, NO$_2$ and O$_2$, adsorbed over eleven different
metal substrates, namely Ag(111), Al(111), Au(111), Co(111), Cu(111), Ir(111), Ni(111), Pd(111),
Pt(111), Sc(111), Y(111). The specific reaction geometries employed for the validation are
provided as Supplementary Information. All the DFT reaction energies retrieved from the
Catalysis-Hub.org database had been computed within the BEEF-vdW approximation to electronic exchange and correlation~\cite{Wellendorff2012}.

The parity plots resulting from this comparison are shown in Figure~\ref{fig:mbe_validation}. The error bars present in the
plots are computed using standard error propagation based on the errors obtained from the OLS regressions.

From Figure~\ref{fig:mbe_validation}a, it is clear that the model captures to an exceptional level of 
accuracy the variation in BE in these systems: 
the correlation coefficient of the parity plot is 0.93, the MAE is 0.77 eV and the RMSE is 1.02 eV. 
When searching for a molecule with a specific adsorption energy with a typical range of 20 eV the MAE of
0.77eV is sufficient to reduce the number of candidate molecules by well over an order of magnitude.
It is, nevertheless, interesting to note that there are outliers to this general trend.
For example, the model predicts a favourable adsorption energy for several molecules (C$_2$H$_6$,
H$_2$O, NH$_3$ and N$_2$O) on Pd(111) (data provided in the Supplementary Information), whereas the DFT data predicts unfavourable adsorption energies. In all of 
these cases, the BE is in the region of $\pm 1$ eV, which is consistent with the MAE found in the training of the model. We can tentatively conclude from this that the current model is less reliable in describing the dissociative adsorption in this weak bonding region.
Another possible explanation for this behaviour can also be found in the different exchange and 
correlation functional employed in our study: as explained in the Method section, we used the
generalized gradient approximation (GGA) for the MBE calculation, whereas the data retrieved from
Catalysis-Hub.org were all computed within the BEEF-vdW approximation, which has specific corrections
to take into account the dispersion interactions. The absence of the latter in our model can explain
the differences arising from the weakly bonded systems.

Apart from these outliers, the prediction of tabulated values is remarkably accurate. As for the original fit to the training set, the intercept of the model has a significant associated error and P-value. It is therefore interesting to test the performance of the model when this parameter is neglected, this data is displayed in Figure~\ref{fig:mbe_validation}b. It is notable that i) without $a$ there is only a slight offset in the predicted BE which does not affect the general behaviour of the parity plot, with $R^2$, MAE and RMSE essentially unchanged, and ii) the estimated error bars for each energy are substantially reduced. This latter observation is
explained by the fact that half of the error in the predicted BE is due to the uncertainty of $a$, which is the coefficient with the highest relative error and P-value.

The discussion above demonstrates that the current model provides a low cost prediction of the BE to homogeneous metal substrates. It is interesting to speculate on the extension of the model to a more general framework for predicting adsorption to a wider range of substrates. A natural extension would be to design a simple MBE cluster calculation for the oxide and carbonate substrates, which are essential in many technological applications and for which there is currently a lack of predictive models regarding molecular adsorption.

\subsection*{Comparison with SISSO approach}

A possible concern in using the proposed approach is that our model only uses OLS regressions,
which could appear simplistic compared to the current state-of-the-art approaches. In particular, one of the most advanced
methods in extracting effective descriptors to predict materials properties is the so-called Sure Independence Screening
and Sparsifying Operator (SISSO) algorithm~\cite{Ouyang2018}. SISSO can identify the best descriptors among a set of
physical properties, determining the optimal subset. Moreover, it can also identify the most accurate
mathematical expressions to obtain the optimised relationship for the available data. It is, therefore, natural
to benchmark the results shown in the previous sections with the ones obtained by employing the SISSO algorithm.

The first step is to identify whether our approach provides an accurate set of descriptors and
identifies the proper mathematical expression. Therefore, alongside the three descriptors employed in the previous
sections, we added eight additional physical and chemical properties of the molecule and the metallic substrate to
increase the set of descriptors, among which SISSO would determine the optimal ones. Namely, we choose the number of
valance electrons of the atomic specie in the metallic substrate ($\text{N}_{\text{ve}}$), the surface energy of the metal
($\gamma_{\text{met}}$), the first ionization potential of the metal ($\text{I}_{1}$), the volume of a single atom in the
metallic substrate ($\text{V}_{\text{met}}$), the metal electronegativity ($\chi_{\text{met}}$), the HOMO
($\text{HOMO}_{\text{mol}}$), LUMO ($\text{LUMO}_{\text{mol}}$) and molar mass ($\text{M}_{\text{mol}}$) of the
molecule as additional physical/chemical descriptors to process in the SISSO algorithm. These data have been gathered
as tabulated values (namely, $\text{N}_{\text{ve}}$, $\text{I}_{1}$, $\text{V}_{\text{met}}$, $\chi_{\text{met}}$ and
$\text{M}_{\text{mol}}$), by DFT values provided by the Materials Project database ($\gamma_{\text{met}}$)~\cite{Jain2013}
and by performing DFT calculations ($\text{HOMO}_{\text{mol}}$ and $\text{LUMO}_{\text{mol}}$) as detailed in the Method
section.

Once we applied the SISSO algorithm to the training data shown in Fig.~\ref{fig:mbe_large}, we obtained the best fitting
with three coefficients by using the following equation (the data obtained by the SISSO algorithm are provided as
Supplementary Information):

\begin{equation}
    \begin{split}
    \text{BE}_{\text{SISSO}} &= 0.787 - 0.044 \times \text{V}_{\text{met}} + \\
    &+ 0.178 \times \left( \text{W}_{\text{F}} - \frac{ \text{E}_{\text{gap}}}{2} \right) + 1.010 \times \text{MBE}
    \end{split}
    \label{eq:sisso}
\end{equation}
\\
with an associated RMSE of 0.65 eV. The best descriptors identified by the SISSO algorithm were the MBE and the difference
between the substrate work function and half of the HOMO-LUMO gap, as already identify in with our approach,
alongside $\text{V}_{\text{met}}$. The former is the only different descriptor compared to our model, in which we used
the cohesive bulk energy of the substrate. Most notably, Eq.~\ref{eq:sisso} shows a linear relationship
with similar coefficients to the OLS fitting shown in Table~\ref{tab:regr_coeff-c}, and we noticed a reduction of the
RMSE of around 5\%. Therefore, our proposed model shows similar results both in identifying the best
descriptors and in the RMSE, thus the SISSO methodology validates our approach. The significant difference in one of the descriptors identified
by the fitting should not be seen as a downside of our work, rather as a different way to interpret the
problem of predicting BE: we have identified the descriptors in our methodology by finding physical and
chemical properties adequate to the studied systems, whereas the SISSO algorithm is looking only to the
primary features that best fit mathematically the BE data.

\section*{\label{sec:conclusion}Conclusion}

In summary, we report a new model of molecule-surface binding based on the combination of \textit{ab initio} calculations with machine
learning algorithms, such as ordinary least squares regression. This model provides in minutes and with limited computational effort a reasonable estimate of the adsorption of small
molecules to metal substrates given a set of easily computed descriptors. The model distinguishes different reaction sites and between molecular and
dissociative adsorption accurately, especially for larger adsorption energies (values greater than $\pm 1$
eV). Compared with an independent and well established database of computed adsorption energies, the predicted values suggest that the model is transferable in that it can provide equally accurate BE predictions for a variety of functional groups and surfaces from outside the training set. We have benchmarked our model against the SISSO algorithm,
finding that the best fitting of the BE data is with a linear equation, and our choice of the descriptors is
very close to the best-case scenario identified by the algorithm, with a difference in the RMSE of around 5\%.

The model is constructed so that its extension to different substrates (e.g., oxides
and carbonates) and technically relevant functional groups is straightforward. 
We expect the model to find widespread use in a variety of applications. For example, the innovation of new coatings for friction and corrosion reduction and the development of novel anti-pathogen coatings to reduce disease transmission via surfaces. 

\section*{Acknowledgements}

This work made use of the high performance computing facilities of Imperial College London. The authors
thankfully acknowledge the funding and technical support from BP through the BP International Centre
for Advanced Materials (BP-ICAM). The authors want also to thank Dr. Giuseppe Mallia for the fruitful 
discussions. The pictures present in this article are generated thanks to
XCrySDen~\cite{Kokalj1999,Kokalj2003} and Matplotlib~\cite{Hunter2007}.

\bibliography{arxiv_main}% Produces the bibliography via BibTeX.

\end{document}